\documentclass[12pt,preprint]{aastex}
\usepackage{graphicx}

\title{Constraints on Black Hole Jet Models Used As Diagnostic Tools of Event Horizon Telescope Observations of M87}
\author{Brian Punsly\altaffilmark{1}}
 \altaffiltext{1}{1415 Granvia Altamira, Palos Verdes Estates CA, USA
90274: ICRANet, Piazza della Repubblica 10 Pescara 65100, Italy and
ICRA, Physics Department, University La Sapienza, Roma, Italy,
brian.punsly@cox.net}
\begin{document}

\begin{abstract}
Jet models of Event Horizon Telescope (EHT) data should also conform to the observed jet profiles just downstream. This study evaluates conformance of models of black hole jets to images of the innermost jet of M87. This is a basic test that should be passed before using them to perform a physical interpretation of EHT data. Recent 86 GHz Very Long Baseline Interferometry observations of M87 have revealed the morphology and size of the jet near its source ($<65$\,M, or 0.06 lt-yrs after correcting for line of sight to the jet, where $M$ is the black hole mass in geometrized units) for the first time. Current transverse resolution indicates that this region is dominated by flux emanating from the edge of the jet. The observed inner jet profiles are compared to all existing published synthetic radio images constructed from ``state of the art" 3-D numerical simulations of the black hole accretion system in M87. Despite efforts to produce the characteristic wide, edge dominated jet, these models are too narrow (by a factor of $\sim2$) in the region 0.06 - 0.32 lt-yrs from the source, even though the jets (spine and/or sheath) in the image plane might appear conformant farther downstream. Furthermore, the synthetic radio images are not edge dominated 0.06 - 0.32 lt-yrs from the source, but spine dominated. Analyses that implement these models as physical diagnostics of EHT visibility amplitudes are therefore suspect. Thus, these inner jet characteristics are important considerations before applying simulations to the EHT data.
\end{abstract}

\section{Introduction}
The study of relativistic jets from active galactic nuclei (AGN) depends on the imaging provided by large scale radio and microwave interferometers. These instruments provide the only high resolution images. Most of our theories and insight have been based on early images from 25 to 40 years ago. These observations image the jet very far from the source in terms of a scale set by the size of the central supermassive black hole (BH). Namely, if the mass of the central BH is $M$ in geometrized units then these ``high resolution"observations were defining features $10^{3}$M -$10^{6}$M from the central BH. One of the great advances in modern astronomy is the improvement in resolution of these interferometers in the past decade. In particular, millimeter band very long baseline interferometry (VLBI) routinely produces quality images at 7 mm. However, it is the recent ability to produce robust images at 3 mm that has finally resolved the region $\sim 60-350$ M from the central BH in M87, after line of sight (LOS) de-projection (distances are de-projected unless otherwise stated). Thus, we are finally detecting size and morphology that can be reliably related directly back to the source itself as opposed to the huge extrapolations from light years away. The next decade promises refinement of these images with the addition of more baselines to the VLBI network with grand goal of quality images at 0.8 mm and 1.3 mm with the Event Horizon Telescope (EHT). The physical interpretation of the EHT images will depend on the suite of numerical models that they are tested against. This study takes the first critical look at whether the models that we currently have are suitably conformant to the existing data $\sim 60-350$ M from the central BH to justify extrapolation back to the EHT emission region. This is not a minor detail of the theory, this is the region that has the strongest direct causal connection to the jet launching mechanism.
\par In spite of the early stages of these observations, the 86 GHz VLBI observations have already provided robust results that could not have been expected from simple jet models extrapolated towards the source from $10^{3}$M -$10^{6}$M away. The nearby, $<350$M, size and morphology is directly compared with computations produced from existing theories and numerical models of BH jets for the first time. Previous treatments have stressed conformance to larger scales, without rigorous quantitative analysis at small distance near the jet source, the region of primary interest.
\par It is not trivial to compare the profile indicated by 86 GHz VLBI observations of the inner jet of M87 to existing theory and models of BH driven jets. The jet boundary in simulations has various definitions \citep{dex12,mck12,sad13}. A particular notion of jet``width" is undefined in the image plane and cannot be compared to observation without an emissivity profile convolved with the restoring beam of comparison radio images. The vast majority of simulations in the literature are lacking in this regard. Fortunately, there are two numerical studies that have produced synthetic radio images of their numerical models, so that a direct comparison with the VLBI data is now possible \citep{mos16,cha18}. The LOS that is used to project these models onto the sky plane is crucial. The 86 GHz VLBI data of \citet{kim16} is processed assuming a LOS of $18^{\circ}$, while the models of \citet{mos16,cha18} assume $20^{\circ}$ and $17^{\circ}$, respectively. In this analysis we assume a range of LOS, $17^{\circ}$ to $25^{\circ}$ \citep{sta06,mer16}. Furthermore, it is assumed that the mass of central BH is $M_{bh} \approx 6 \times 10^{9}M_{\odot}$ or $M \approx 8.86 \times 10^{14} \rm{cm}$ in geometrized units, which equates to $\approx 3.5\mu\rm{as}$ at 16.7 Mpc \citep{geb11}. Section 2 will describe the inner jet profile as given by the 86 GHz VLBI observations. Section 3 compares and contrasts VLBI images with numerical models.

\begin{figure}
\begin{center}
\includegraphics[width= 0.80\textwidth]{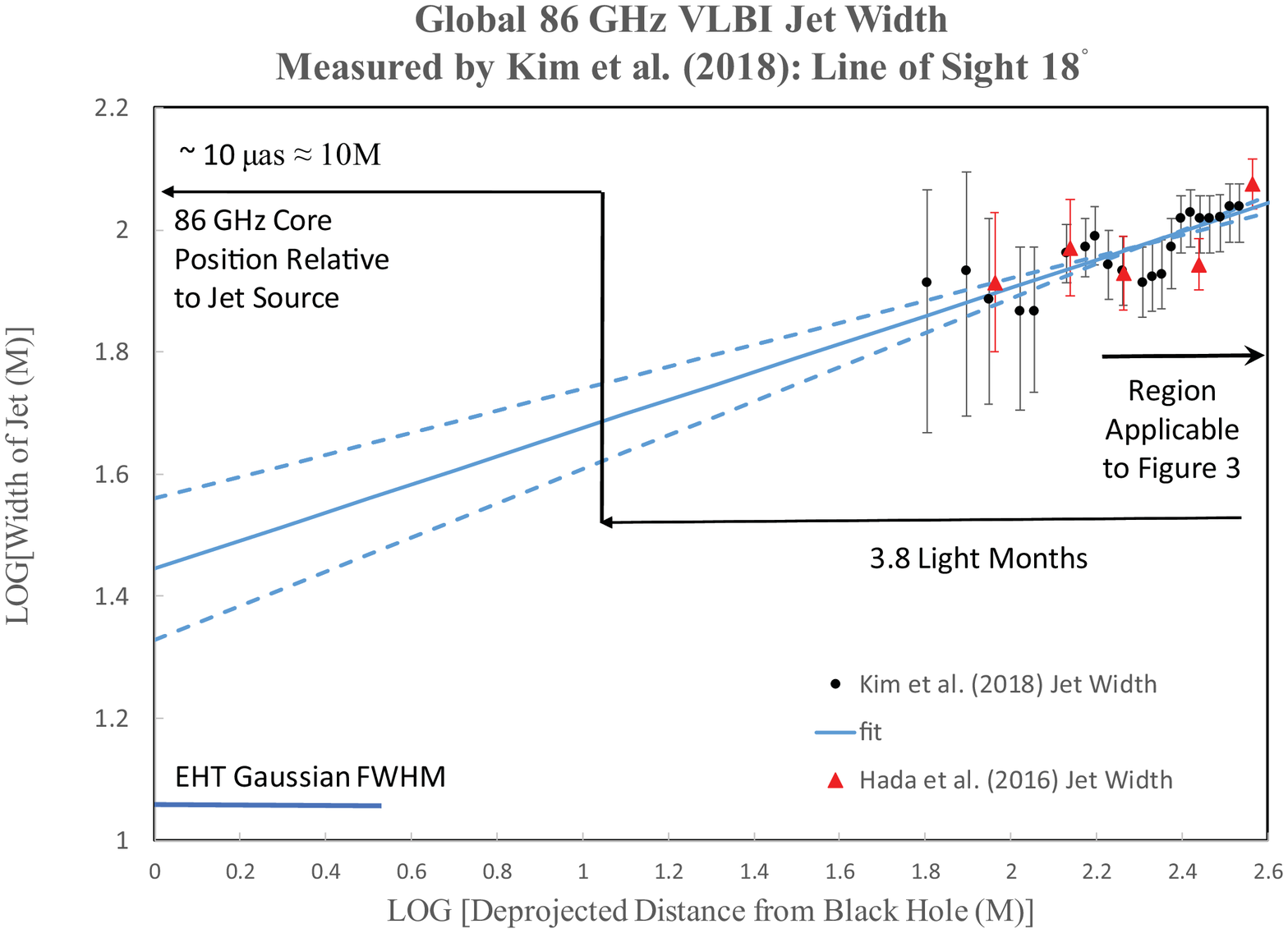}
\includegraphics[width= 0.80\textwidth]{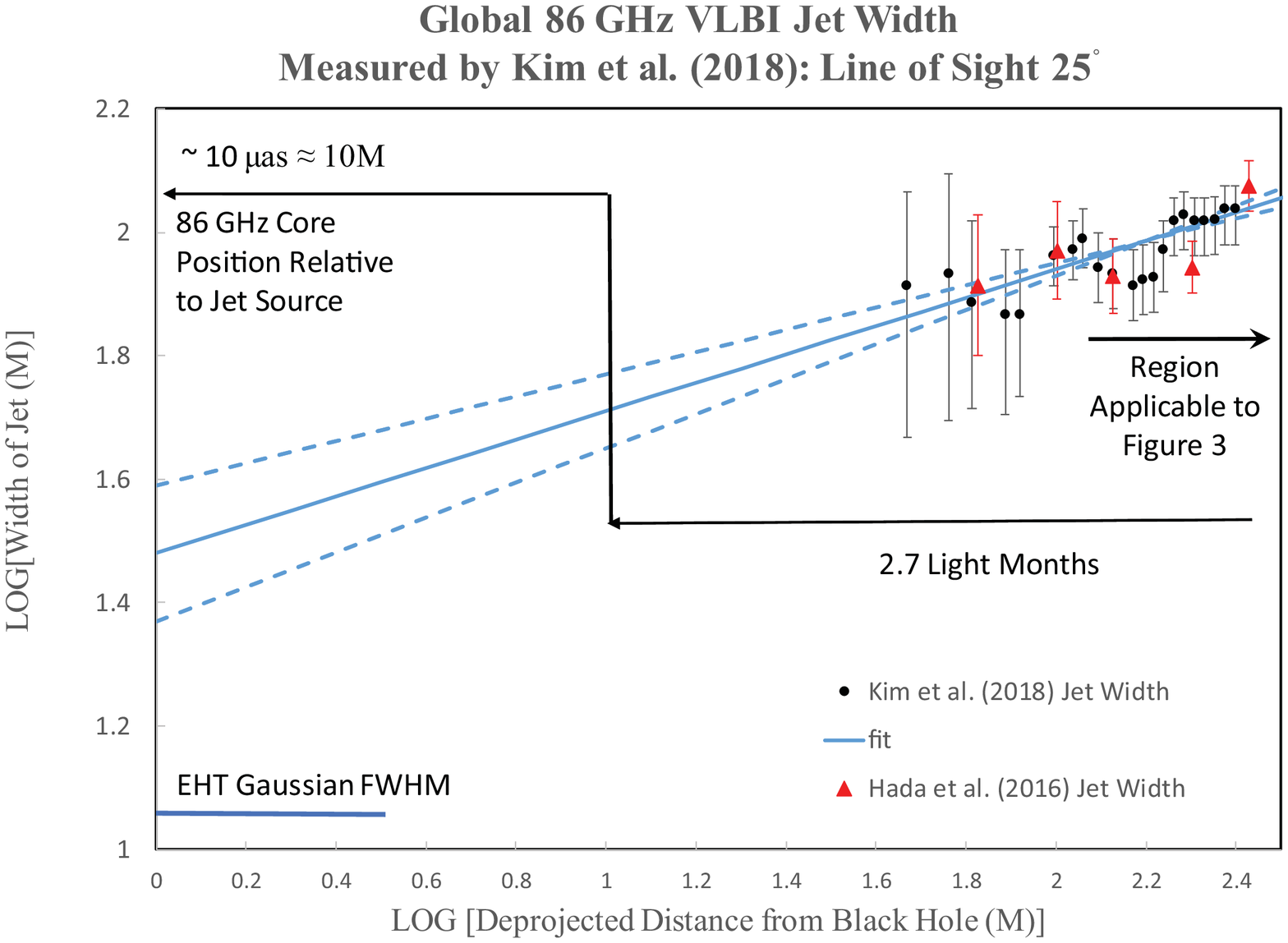}
\caption{The top panel shows the fit with uncertainty in both variables to the M87 86 GHz VLBI jet width profile. The data is from of \citet{kim16} who assume a LOS of $18^{\circ}$. The fitted data extends $\sim 3.8$ light months from the unresolved radio core. The bottom panel is the fit assuming a LOS of $25^{\circ}$. The powerlaw index is unchanged, but the width extrapolates to be wider at $z=0$. The location of the EHT correlated flux is not known, but it is shown schematically in the lower left hand corner.}
\end{center}
\end{figure}
\section{The Innermost Profile of the Jet of M87}

M87 has been monitored with the Very Long Baseline Array (VLBA) at 43 GHz since the beginning of the century \citep{wal16}. This has provided detailed information on scales larger than 0.3 mas. In this paper, longer baseline data from the Global mm-VLBI Array (GMVA) at 86 GHz is used to study the jet structure on smaller scales,
0.05 - 0.35 mas \citep{kim16}.
\par The top frame of Figure 1 shows the jet width as derived by \citet{kim16} from the GVMA observations that they studied. These data are extracted from the stacking of 5 observations spread out from 2004 to 2015 and was presented in their Figure 6. The analysis is facilitated by the existence of two pronounced ridges of emission that has been referred to as edge brightening. The jet width is the distance between the peak brightness of the ridges. The \citet{kim16} data in the plot extends 340 M from the jet source. The bottom frame changes the assumed LOS from $18^{\circ}$ to $25^{\circ}$. These plots are restricted to a smaller distance from the source than was chosen in \citet{kim16} during their fitting process. By choosing many points farther from the BH, the fit to the jet in the vicinity of the source is not optimized since it is overwhelmed by weighting the least squares fit with residuals of the numerous points $\sim 1000$ M from the source. Thus, most of the points that are displayed in Figure 1 are circumvented by their fit that is dominated by points farther out. This might be justifiable if the one insisted on a single powerlaw fit over all distance scales. However, this is not directly motivated by the observational data itself, but by simple theoretical models \citep{bla79}. The \citet{kim16} powerlaw fit to the jet width, $W(z)$, as function of axial displacement along the jet, $z$, is $W(z) \propto z^{-k}$, $k=0.498\pm 0.025$. This study is an attempt to be a bit more rigorous. A uniform powerlaw over the innermost few light years of jet propagation is not assumed in the following. The data is restricted to a range of 3.8 light months ($z<340$M, 100M$\approx$ 0.0937 lt-yr.). There are certainly ample data in this regime and it is far from the source in terms of the FWHM of the circular Gaussian fit to the EHT emission at 230 GHz ($\sim 40 \mu$as) shown schematically near the left edge of both panels of Figure 1 \citep{doe12,aki15}. Secondly, the uncertainty in the data is considered in the fit to $W(z)$. The uncertainty in $z$ is half the distance between the points in Figure 6 of \citet{kim16} and the error in $W(z)$ is from Figure 6 of that paper. The method of \citet{ree89} is used to fit the data with uncertainty in both variables. The powerlaw index is $k=0.230\pm 0.049$. The \citet{kim16} data is validated by the close alignment of the 2014 VLBI data of \citet{had16} in Figure 1. The dashed lines in Figure 1 indicate the standard error of the \citet{ree89} fit to the data. Note that the synchrotron self-absorption core shift at 86 GHz is assumed to be 10.5 $\mu$as based on the analysis of \citet{had11}. The fits are insensitive to the uncertainty in the small core offset, so it is not critical to know this value exactly.
\par The fit in Figure 1 has two interesting implications. The jet within 3.8 light months from the source is much more collimated than a parabolic jet, $k=0.5$. Secondly if one extrapolates to within $\sim 20$M of the central BH, we see that the jet is $\sim 2-3$ wider than the size of the circular Gaussian fit to the correlated EHT flux. Thus, this extrapolation is not justified. The jet must have a very large opening angle at small $z$ in order for the EHT data to join with the powerlaw fit to the GMVA data 0.7- 3.8 light months of the source. The maximum $z$ at which this large opening can collimate is the innermost GVMA data point. This results in the minimum possible intrinsic opening angle of the jet base, $>65^{\circ}$, where the fitted powerlaw was chosen to represent the innermost data point. It is larger if we use the data point itself, extrapolate the fit closer to the BH, or assume a LOS $>18^{\circ}$.
\section{Comparison of the Inner Jet Profile of M87 with Models of Black Hole Jets} Only 3-D models of radiatively inefficient BH accretion are considered relevant for M87 \citep{nar94}. The nexus to VLBI observation requires a synthetic image to be made from the simulation with the same restoring beam (same resolution limits) as the comparison radio image. Currently, there are only two numerical studies for which such images have been generated by the researchers. The magnetic flux in the central funnel of the accretion disk ranges from modest to large, SANE and MAD (magnetically arrested), respectively \citep{igu03,tch11,nar12}. The SANE simulation radiates entirely from a funnel wall jet (FWJ) at the funnel/disk interface \citep{mos16}. The MAD simulations in \citet{cha18} have been argued to be required to power the jet in M87. They radiate from the FWJ and, to some degree, the adjacent regions of the interior jet. Both models assume rapidly rotating BHs, $a/M = 0.9375$, where $a$ is the angular momentum per unit mass of the BH.
\subsection{Funnel Wall Jets} In Figure 6 of \citet{mos16}, the simulated emissivity is restored with the same beam as the stacked 43 GHz VlBI observations of \citet{had13}, 0.30 mas x 0.14 mas that is shown in the top frame of Figure 2. Figure 2 provides some metrics to quantify the jet profiles within the context of the radio (real and synthetic) images. Fortunately, there are contour levels in their Figure 6 that can be used to evaluate the observed surface brightness distribution, $S(\rho, z)$. The axial displacement along the red jet center-line 0.2 -0.4 mas from the radio core, $z$, and transverse displacement, $\rho$, as projected on the sky plane are illustrated in the top frame. Using this coordinate system, Figure 2 explores the axial surface brightness, $S^{A}(z)\equiv S(\rho=0, z)/S(\rho=0, z=0)$ and the jet 0.5 contour or half width,
\begin{eqnarray}
&& S_{0.5}(z) =\rho_{+0.5}(z) - \rho_{-0.5}(z), \; S[\rho=\rho_{+0.5}(z), z]\Theta(\rho)\equiv 0.5S(\rho=0, z), \nonumber\\
&& \rm{and} \; S[\rho=\rho_{-0.5}(z), z]\Theta(-\rho)\equiv 0.5S(\rho=0, z)\; ,
\end{eqnarray}
where $\Theta$ is the Heaviside step function.
One can determine $S^{A}(z)$ and $S_{0.5}(z)$ everywhere the centerline crosses one of the plotted surface brightness contours. The contour levels have a ratio of $\sqrt{2}$. The top frame of Figure 2, displays the evaluation of $S_{0.5}(z)$ at z =0.26 mas by the length of the blue segment orthogonal to the center line. The bottom frame of Figure 2 shows that $S^{A}(z)$ in the model decays slightly slower than the radio image as z increases. Yet, $S(\rho, z=\rm{contanst})$ decays much more rapidly with $\rho$ in the model than the radio image. In order compute the uncertainty of the $S_{0.5}(z)$ estimate, note that the monotonic increase in $S_{0.5}(z)$ tracks the increase in the double ridge line estimate of $W(z)$ in Figure 3 of \citet{had13} with a positive offset $\approx 0.2\, \rm{mas}$. Thus, the systematic uncertainty of the $S_{0.5}(z)$ estimate, $\sigma_{s}$, is identified with the uncertainty of $W(z)$ in \citet{had13}, Figure 3, as 12\%. The uncertainty in the contour intensity level from thermal noise is chosen as 3 rms = 3.9 mJy/beam. This induces an uncertainty in the position of the contour level in the image plane. For example, if the intensity level is increased by 10\%, since the contour levels are every factor of 1.41, this increase induces a shift of approximately $0.1/0.41$ towards the adjacent contour level along the blue line in the top frame of Figure 2. This shift of the intensity contour due to thermal noise is the uncertainty, $\sigma_{t}$. The resultant uncertainty of $S_{0.5}(z)$ is $\sigma =\sqrt{\sigma_{s}^{2} + \sigma_{t}^{2}}$. The axial positional uncertainty is 1-10\% in \citet{had13}. This motivates the conservative choice of uncertainty, the smaller of: half the distance to the adjacent contour and $0.1z$.
\par Even though, the intrinsic emissivity distribution in \citet{mos16} is approximately a $\delta$-function at the maximum value of $\rho$ at each $z$, the image appears spine dominated, not edge dominated because of the ``large" 0.3 mas beam width in the transverse direction. By contrast, the VLBI image is not spine dominated, but edge dominated for $z>0.25$~mas. Thus, these FWJ simulations do not represent the M87 jet near its source.
\begin{figure}
\begin{center}
\includegraphics[width= 0.75\textwidth]{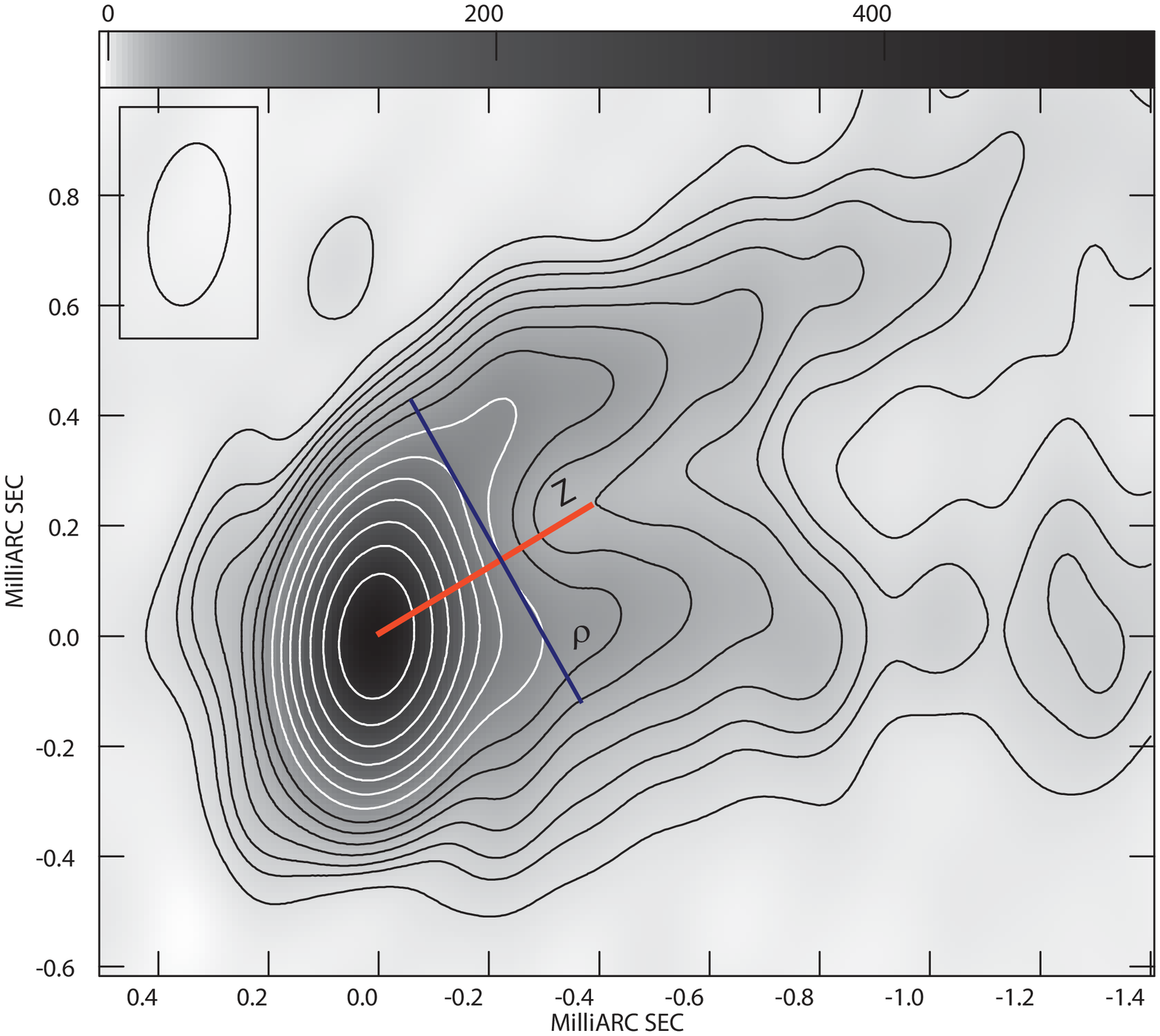}
\includegraphics[width= 0.75\textwidth]{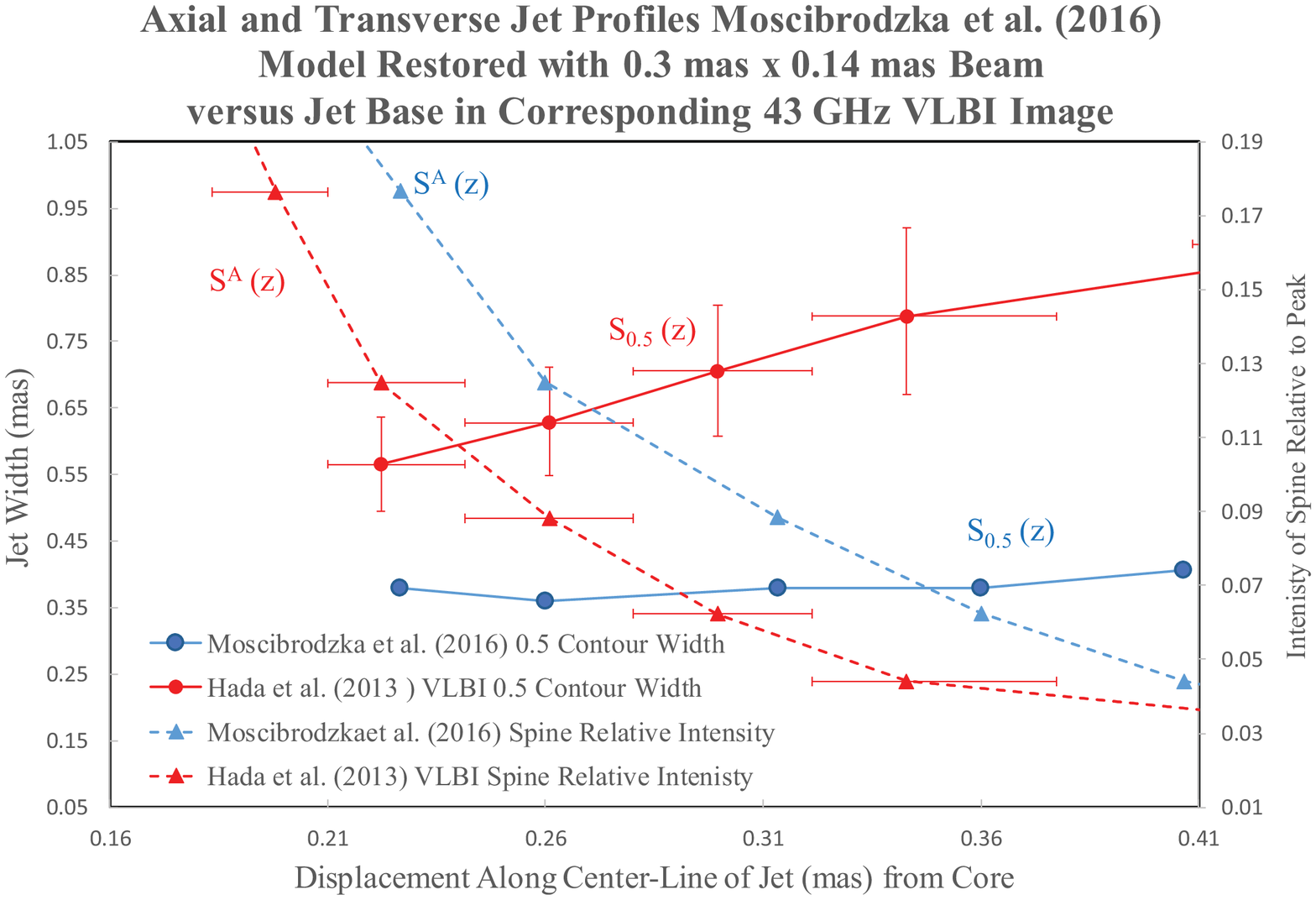}
\caption{The top frame shows the 43 GHz VLBA image provided by K. Hada that is used for comparison with the FWJ model. The blue $\rho$-axis crosses the red z-axis at $z= 0.26$ mas. The contours are $\sqrt{2}$ increments. The length of the blue segment indicates that $S_{0.5}(z= 0.26\, \rm{mas}) = 0.63$ mas. The bottom frame compares the FWJ model surface brightness with the image after restoration with the same beam. The lines between points do not represent intermediate values, but merely draw one's eye to the overall trend.}
\end{center}
\end{figure}
\par It is of interest to quantify the intrinsic difference between the width of the inner jet of M87 and the width of the FWJ. This is possible due to the fact that the luminosity distribution is restricted to a very thin boundary layer in both M87 and the model. At 0.2 mas from the source, it was shown in Figure 5 of \citet{pun18} that the \emph{intrinsic} surface brightness associated with the 86 GHz VLBI observation of \citet{had16} is approximately a double $\delta$-function, $S_{\rm{int}}(\rho, \, z = 0.2\rm{mas}) \approx S_{\rm{Hada}}[ 0.48\delta(\rho + 0.15 \rm{mas}) + \delta(\rho - 0.15 \rm{mas})]$, where $S_{\rm{Hada}}=19 \, \rm{mJy/beam}$ is a normalization constant. The observed $S(\rho, \, z = 0.2\rm{mas})$ restored with the full beam is highly blurred, but still edge dominated \citep{had16}. Similarly, from Figure 4 of \citet{mos16}, $S_{\rm{int}}(\rho, \, z = 0.2\rm{mas}) \approx S_{-} \delta(\rho + 0.075 \rm{mas}) + S_{+}\delta(\rho -0.075 \rm{mas})$, where $S_{+}$ and $S_{-}$  are constants that cannot be determined from the article. The blurring of the surface brightness by the VLBI restoring beam reduces the ratio of the 0.5 contour widths in the bottom frame of Figure 2 relative to the intrinsic ratio of the physical dimensions, given by: $\approx$ [2(0.15 mas)]/[2(0.075 mas)] = 0.3mas/0.15mas =2.
\subsection{MAD Models} Figure 3 compares the width of the inner jet, as determined from observation, with numerical models in \citet{cha18}, H10 and R17. The simulations of \citet{cha18} match the $55^{\circ}$ opening angle in the image plane gleaned from 43 GHz VLBA \citep{wal18}. But this agreement with the jet occurs at $z =0.5\, -\,1.9$~mas. By contrast, Figure 3 considers $z < 0.4$~mas (recall the different value of $k$ found by \citet{kim16} by considering more distant points than are chosen in Figures 1). Figures 10 and 11 of \citet{cha18} are an effort to directly compare the simulations to VLBI images at 43 GHz and 86 GHz, respectively. The authors have processed the observational data and the numerical results with the same restoring beam. In spite of this, the comparison is not trivial because the jet morphology is different between the models and the observations within 0.4 mas. The flux from the M87 jet is dominated by the edges on these small scales (see Figure 5 of \citet{pun18}), but the images produced by the simulation have jets in which the flux is concentrated along the axis of the jet. One might say that the M87 jet is extremely edge brightened for $z<0.4$~mas and the H10 and R17 simulations are edge darkened for $z<0.4$~mas.
\begin{figure}
\begin{center}
\includegraphics[width= 0.8\textwidth]{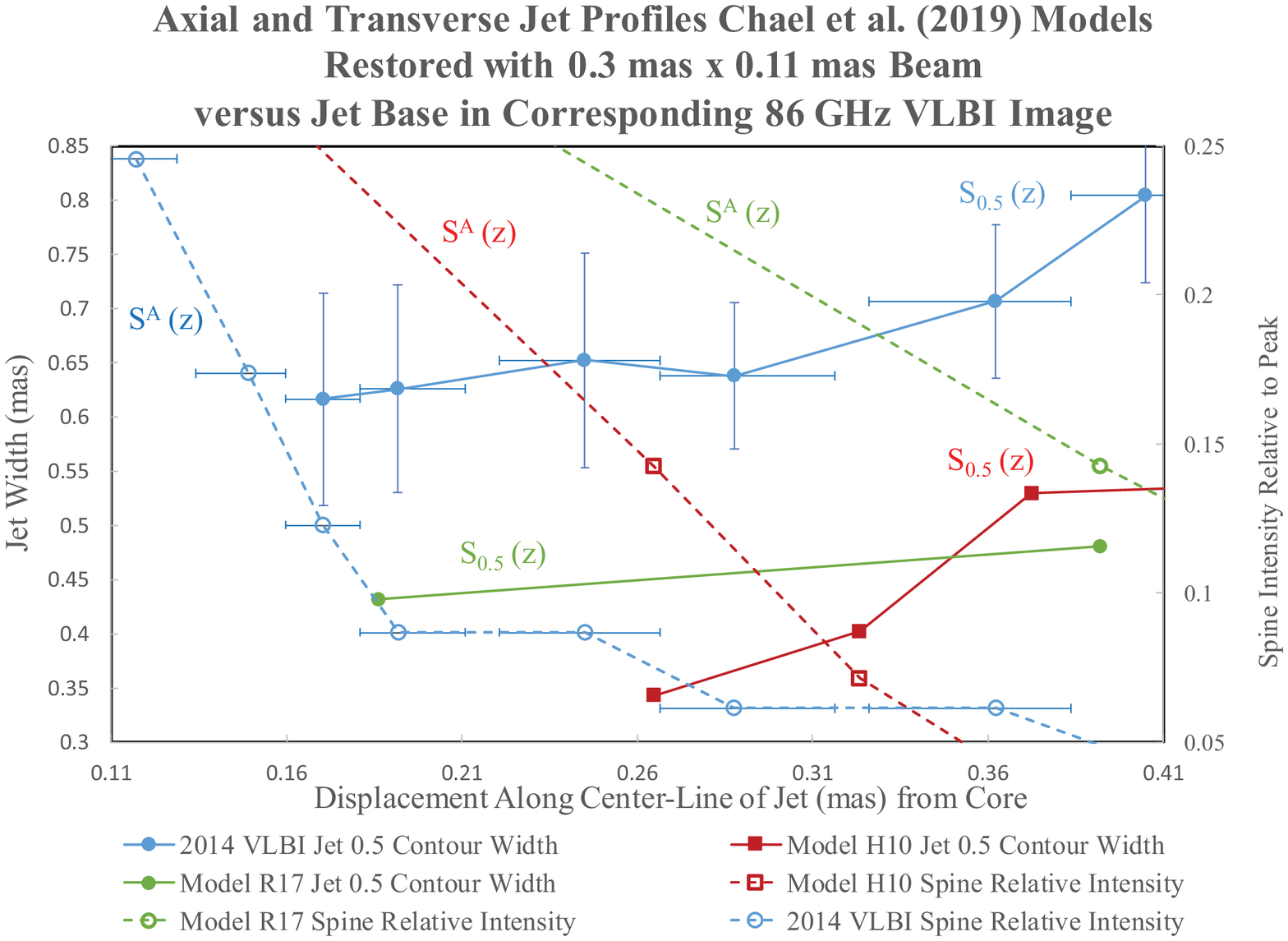}
\includegraphics[width= 0.8\textwidth]{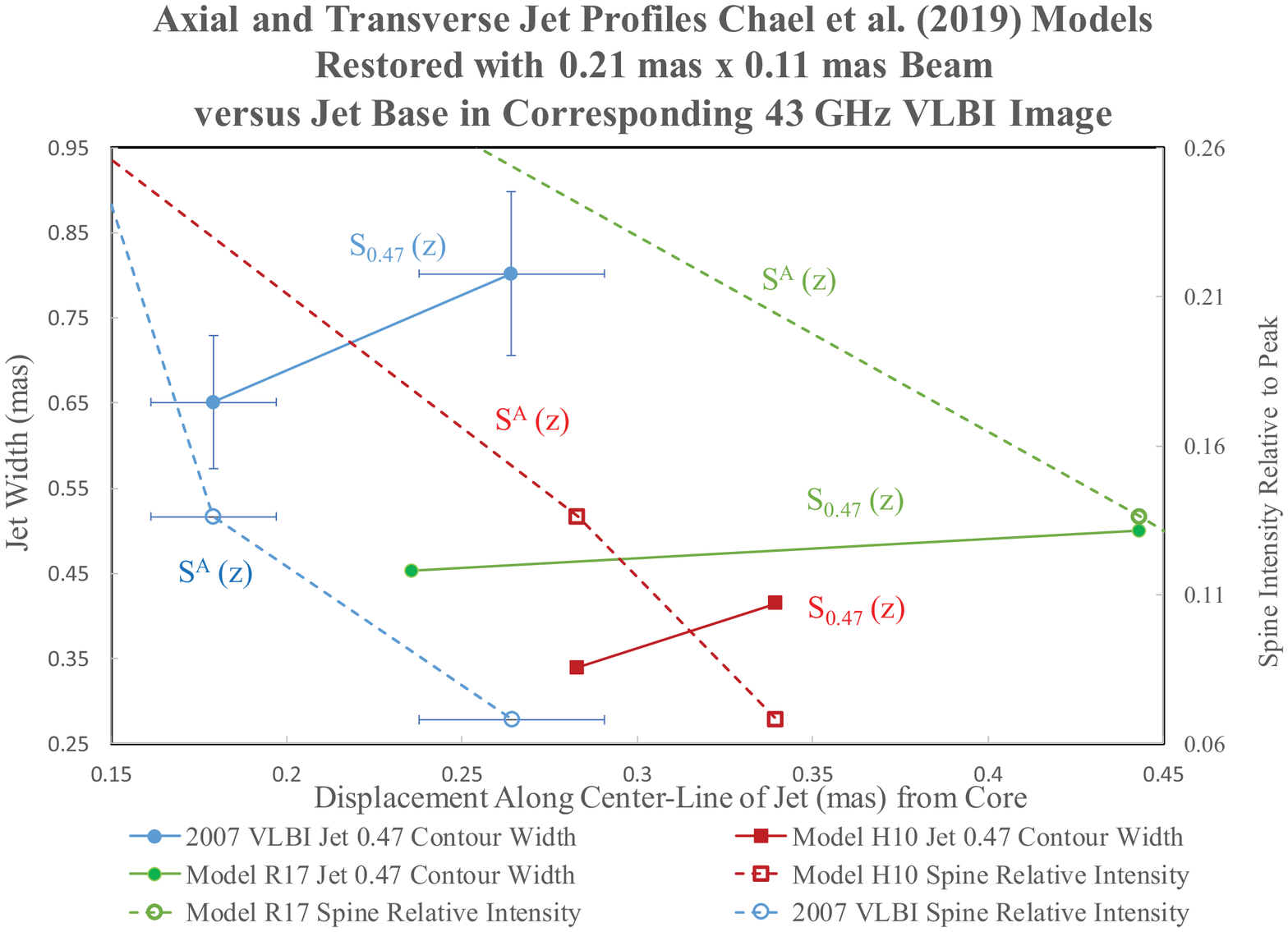}
\caption{The top frame compares the inner jet profile of the \citet{had16} 86 GHz observation with the two models H10 and R17 of \citet{cha18} after restoration with the same beam. The bottom frame is the same at 43 GHz. See the text for details.}
\end{center}
\end{figure}
\par The top frame of Figure 3 applies the metrics used to describe $S(\rho, z)$ for the FWJ, to the 86 GHz image and models presented in \citep{cha18}. Since the 86 GHz image was originally restored in \citet{had16} with a similar beam, the original image was used since it is has higher contour resolution (every $\sqrt{2}$ instead of 2). Figure 9, and Table 1 of \citet{had16} provide the data for $\sigma_{s}$ and $\sigma_{t}$, respectively that are used in computing the errors (as in Figure 2) in the top frame of Figure 3, see also the demarcated region in the top right hand corner of Figure 1. At 43 GHz, in the bottom frame, the contour levels in \citet{cha18} are at 47\% not 50\% (the contour decrement is 2.15 not 2) so one must use 0.47 instead of 0.5 in Equation (1).The same parameters are used to compute the uncertainty in the bottom frame as were used in Figure 2. At both frequencies $S(\rho, z=\rm{contanst})$ decays much more rapidly with $\rho$ in the models compared to the observation and $S^{A}(z)$ decays much more slowly with $z$ in the models compared to the observations. The \citet{cha18} simulations are much narrower than the observed jet in M87 at 43 and 86 GHz for $z<0.4$~mas. Furthermore, the spine is too pronounced, M87 is very edge dominated at 86 GHz for $z<0.4$~mas. Figure 3 indicates that these MAD simulations do not represent the M87 jet accurately near its source.

\section{Conclusion} This study proposes guidelines based on observation that models should follow before application to the EHT region of M87. Namely, the observed jet profile within 0.35 mas of the core in M87 is compared with jet profiles from models of BH driven jets. In Section 2, it was shown that the jet base within 0.06 lt-yrs of the source is widely flared with an intrinsic opening angle $>65^{\circ}$ and the jet is highly collimated for 0.06 lt-yrs $<z<$ 0.32 lt-yrs. In Section 3, it was demonstrated that all published synthetic radio images from BH jet models produce jets that are too narrow relative to M87 on scales 0.06 lt-yrs $<z<$ 0.32 lt-yrs. The synthetic images are not edge dominated, like the images of M87, but spine dominated for 0.06 lt-yrs $<z<$ 0.32 lt-yrs. These are not minor discrepancies of the theory, this is the region that has the strongest direct causal connection to the jet launching mechanism.
\par Three models were analyzed with different electron temperature prescriptions (although all produce a funnel and funnel wall hotter than the disk) and different funnel magnetic field strengths. All showed a similar discrepancy in the axial and transverse intensity. The discrepancy that was found is not an issue of time variability, it occurred with different telescopes and different frequencies over a $\sim 10$ year period ($t>10^{4}M$). Model parameters such as the treatment of the polar axis, disk size and tilt, funnel magnetization and the jet launching region can affect the jet width \citep{dib12,sad13,pun17}. {It is not clear, nor claimed, that Figures 2 and 3 capture the widest, most edge dominated possible simulated jets for $z<0.35$~mas. It is difficult to assess the simulations without synthetic images. However, a $\pm 20\%$ variation in SANE jet width at $z= 50M$ in code validation studies with convergent numerical resolution is noted \citep{por19}. A $\pm 20\%$ uncertainty, after beam convolution, will not alter the results of Figure 2. There is no such study for MAD jets \citep{por19}. MAD jets spread to wider opening angles due to enhanced internal magnetic pressure. The horizon magnetization, normalized to the accretion rate, is defined in \citet{cha18}, $\phi_{BH} = 55-63$.  These simulations are at or near full magnetic flux saturation in the funnel \citep{tch15}. Therefore, these jets might attain near maximal MAD jet width. More simulated synthetic images are required to explore the generality of these results and the physical assumptions that ameliorate these discrepancies.
\begin{acknowledgements}I would like to thank Kazuhiro Hada for sharing his knowledge of the VLBI data. Andrew Chael and Monika Mo\'{s}cibrodzka graciously shared the details of their synthetic radio images. The Very Long Baseline Array and the High Sensitivity Array (Very Long Baseline Array and the Green Bank Telescope) are operated by National Radio Astronomy Observatory, a facility of the National Science Foundation, operated under cooperative agreement by Associated Universities, Inc.: projects BW088G and BH0186. Partial funding for this work was provided by ICRANet. I would also like to thank an anonymous referee whose comments greatly improved the manuscript.
\end{acknowledgements}

\end{document}